# Electron with arbitrary pseudo spins in multilayer graphene

Worasak Prarokijjak[a], Bumned Soodchomshom[a,*]

[a] Department of Physics, Faculty of Science, Kasetsart University, Bangkok 10900, Thailand

*Corresponding author

Email address: Bumned@hotmail.com; fscibns@ku.ac.th (B. Soodchomshom)

**Abstract**

Using the low-energy effective Hamiltonian of the ABC-stacked multilayer graphene, pseudo spin coupling to real orbital angular momentum of electron in multilayer graphene is investigated. We show that electron wave function in N-layer graphene mimics behavior of particle with spin of $\mathrm{N}\times(\hbar/2)$, where N={1, 2, 3…}. It is said that for N>1 the low-energy effective Hamiltonian for ABC-stacked graphene can not be used to describe pseudo spin ½-particle. The wave function of electron in multilayer graphene may behave like fermionic (or bosonic) particle for N being odd (or even). This work proposes a theory of graphene as a host material of electron with arbitrary pseudo spins, tunable by changing number of graphene layers.

## 1. Introduction

Since graphene has been discovered [1], it has become a promising material for nanotechnology. It also has become a wonder material carrying several interesting physics properties that may connect the condensed matter to high energy physics [2]. In graphene, its carriers mimic the behavior of Dirac fermions. This may lead to Klein tunneling observed in graphene p-n junction [3]. Zitterbewegung, trembling motion, could be studied in graphene system [2-4]. Observing atomic collapse in artificial nuclei on graphene was recently reported [5]. This is related to relativistic quantum effect in graphene to cause charged impurities exhibiting resonances due to atomic collapse states [5]. Significant for building quantum computer, fractional quantum Hall Effect which is tunable by electric field in bilayer graphene has been recently observed [6]. Related to electron in graphene behaving like relativistic fermions, the fundamental physical property of such pseudo relativistic particle is needed to be clarified. One of the important topics is physical property of pseudo spin, which is generated by the presence of its two sublattices, A and B, instead of usual (real) spin [7]. In this paper, the property of pseudo spin of electron coupling to the real orbital angular momentum in ABC-stacked multilayer graphene will be investigated. We propose new pseudo spin operators for number of layers greater than one.

## 2. On the pseudo spin ½ electron in monolayer graphene

Monolayer graphene, a two dimensional honeycomb like atomic structure formed by carbon atoms, has two sublattices A and B in a unit cell. The wave function of electron may be described by two-component wave states $\psi_A$ and $\psi_B$ of electrons in the A- and B- sublattices which play the role of the pseudo spin states ie., $\psi_A \approx \psi_\uparrow$ and $\psi_B \approx \psi_\downarrow$, respectively. Graphene yields the low exited energy dispersion similar to two-dimensional Dirac fermion where the speed of light is replaced by the Fermi velocity $v_F \cong 10^6$ m/s. The momentum-energy relation obeys the relativistic like relation $E = \sqrt{(pv_F)^2 + (mv_F^2)^2}$ where E and p are energy and momentum, respectively. The pseudo Dirac mass "m" may be generated due to breaking symmetry of sublattice energy [8]. Since, pseudo spin in graphene is generated by the presence of sublattice states, it was previously understood that pseudo spin generated by the presence of the sublattice states in monolayer graphene may not be associated with angular momentum [9, 10]. Recently, some fundamental physical properties of pseudo spin in monolayer graphene have been investigated [11, 12]. Mecklenburg and Regan [11] has showed that pseudo spin of electron in graphene is associated with real angular momentum, although it is not real spin. This is because the total angular momentum, $J_{z,l} = L_z + S_{z,l}$, which is perpendicular to graphene plane, is conserved i.e.,

$$[\hat{H}_l, \hat{L}_z + \hat{S}_{z,l}] = 0 . \qquad (1)$$

The $\hat{L}_z = \hat{x}\hat{p}_y - \hat{y}\hat{p}_x$ is the angular momentum operator perpendicular to graphene plane where $< \hat{x}, \hat{y} >$ and $< \hat{p}_x, \hat{p}_y >$ are operators of positions and momentum of electron moving in the graphene plane, respectively. The intrinsic pseudo spin of electron in monolayer is found to be of the form [11]

$$\hat{S}_{z,l} = \frac{\hbar}{2}\begin{pmatrix} 1 & 0 \\ 0 & -1 \end{pmatrix} \approx \frac{\hbar}{2}\sigma_z \qquad (2)$$

Here $\sigma_z$ is known as Pauli spin matrix in the z-direction. Since $H_l$ is Hamiltonian acting on pseudo spinor field $\psi = \left(\psi_A, \quad \psi_B\right)^T$ in monolayer graphene, it may be said that effective electron field in monolayer graphene mimics the behavior of “spin ½ particle” because spin operator in eq.(2) yields Eigen value of $\hbar/2$ [11, 12]. The pseudo spin in graphene may be defined as usual Pauli spin operator, given by

$$\hat{S}_l = \hat{S}_{x,l}\hat{i} + \hat{S}_{y,l}\hat{j} + \hat{S}_{z,l}\hat{k} , \qquad (3)$$

where $\hat{S}_{x,l} = (\hbar/2)\sigma_x$ and $\hat{S}_{y,l} = (\hbar/2)\sigma_y$ with $\sigma_{x(y)}$ are Pauli matrices in x(y-) directions. The unit vectors parallel to the x, y and z direction are $\hat{i}$, $\hat{j}$ and $\hat{k}$, respectively. In general, for all spin operators must satisfy “the commutation rule of an angular momentum” [13]. It is found that pseudo spin operator of electron in monolayer graphene defined in eq.(3) obeys this rule to get

$$\left[\hat{S}_{x,l}, \hat{S}_{y,l}\right] = i\hbar\hat{S}_{z,l}, \left[\hat{S}_{y,l}, \hat{S}_{z,l}\right] = i\hbar\hat{S}_{x,l} \text{ and } \left[\hat{S}_{z,l}, \hat{S}_{x,l}\right] = i\hbar\hat{S}_{y,l} . \qquad (4)$$

The pseudo spin only in the z-direction is associated with real angular momentum because there is no orbital angular momentum in the xy plane. It can be said that in-plane pseudo spin may not be considered as a real angular momentum, since

$[\hat{H}_1, \hat{L}_x + \hat{S}_{x,1}] \neq 0$ and $[\hat{H}_1, \hat{L}_y + \hat{S}_{y,1}] \neq 0$ in two-dimensional system, discussed in ref.[12].

In the next section, we will study the pseudo spin operator coupling to orbital angular momentum $\hat{L}_z = \hat{x}\hat{p}_y - \hat{y}\hat{p}_x$ for electron in ABC-stacked multilayer graphene (see Fig.1). We will show that low energy electrons in N-layer graphene may not behave like spin ½ particle for N>1.

### 3. Pseudo spin and its angular momentum property in multilayer graphene

Let us start with the Hamiltonian derived based on the tight-binding model for the ABC-stacked N-layer graphene at the *K* point, as given by [14]

$$\hat{H}_N = \hbar v_F \begin{pmatrix} \vec{\sigma}.\vec{k} & \tau & 0 & 0 & 0 & \cdots & 0 \\ \tau^T & \vec{\sigma}.\vec{k} & \tau & 0 & 0 & 0 & \vdots \\ 0 & \tau^T & \vec{\sigma}.\vec{k} & \ddots & 0 & 0 & 0 \\ 0 & 0 & \ddots & \ddots & \ddots & 0 & 0 \\ 0 & 0 & 0 & \ddots & \vec{\sigma}.\vec{k} & \tau & 0 \\ \vdots & 0 & 0 & 0 & \tau^T & \vec{\sigma}.\vec{k} & \tau \\ 0 & \cdots & 0 & 0 & 0 & \tau^T & \vec{\sigma}.\vec{k} \end{pmatrix}, \tag{5}$$

where $\tau^T = \begin{pmatrix} 0 & \gamma_\perp \\ 0 & 0 \end{pmatrix} / \hbar v_F$ and $\tau = \begin{pmatrix} 0 & 0 \\ \gamma_\perp & 0 \end{pmatrix} / \hbar v_F$, with $t_\perp$ being the interlayer hoping energy [15]. Here, $\hat{\vec{k}} = \hat{k}_x \hat{i} + \hat{k}_y \hat{j}$ is operator of wave vector, and $\vec{\sigma} = \sigma_x \hat{i} + \sigma_y \hat{j} + \sigma_z \hat{k}$. The Hamiltonian above would act on the 2N-component wave function of

$$\psi_N = \begin{pmatrix} \psi_{A,1} & \psi_{B,1} & \psi_{A,2} & \psi_{B,2} & \cdots & \psi_{A,N} & \psi_{B,N} \end{pmatrix}^T . \tag{6}$$

Near the Dirac point, we have $\psi_{B,1} = \psi_{A,2} = \psi_{B,2} = ... = \psi_{A,N} \cong 0$ , the wave functions for electrons may be approximately given as a two-component wave function of the form

$$\psi_N \to \psi_{N,\mathrm{eff}} = \begin{pmatrix} \psi_{A,1} \\ \psi_{B,N} \end{pmatrix} . \tag{7}$$

The two-component low energy effective Hamiltonian reduced form eq. (5), is usually obtained as [14, 16-18]

$$\hat{H}_N \to \hat{H}_{N,eff} = \frac{(\hbar v_F)^N}{(-\gamma_\perp)^{N-1}} \begin{pmatrix} 0 & (\hat{k}_x - i\hat{k}_y)^N \\ (\hat{k}_x + i\hat{k}_y)^N & 0 \end{pmatrix}. \quad (8)$$

This two-component Hamiltonian would act on the two-component pseudo spin state ie., $\hat{H}_{N,eff}\psi_{N,eff} = E\psi_{N,eff}$ where E is the excitation energy of electron.

We next consider the pseudo spin operator $\hat{S}_{z,N}$, which is associated with orbital angular momentum operator $\hat{L}_z = \hat{x}\hat{p}_y - \hat{y}\hat{p}_x$ for the low energy electron in the N-layer ABC-stacked graphene. We may define the total angular momentum for the low energy electron as of the form $\hat{J}_{z,N} = \hat{L}_z + \hat{S}_{z,N}$. The pseudo spin operator for the N-layer graphene must satisfy the conservation of total angular momentum condition of $[\hat{H}_{N,eff}, \hat{J}_{z,N}] = 0$ .We thus calculate pseudo spin operators via this condition ie.,

$$[\hat{H}_{N,eff}, \hat{L}_z] = \frac{(\hbar v_F)^N}{(-\gamma_\perp)^{N-1}} \begin{pmatrix} 0 & N\hbar(\hat{k}_x - i\hat{k}_y)^N \\ -N\hbar(\hat{k}_x + i\hat{k}_y)^N & 0 \end{pmatrix}$$

$$= -[\hat{H}_{N,eff}, \hat{S}_{z,N}],$$

where

$$\hat{S}_{z,N} = N\frac{\hbar}{2}\begin{pmatrix} 1 & 0 \\ 0 & -1 \end{pmatrix} = N\frac{\hbar}{2}\sigma_z. \quad (9)$$

This result is unusual and very surprising, because the Eigen value of pseudo spin in the z-direction carries magnitude of $\hbar/2$ only for N=1 or in case of monolayer graphene. For the case of N={2, 3, 4, 5,…}, the pseudo spin operator yields its Eigen value of $\{\hbar, \frac{3}{2}\hbar, 2\hbar, \frac{5}{2}\hbar, \ldots\}$, respectively. It is said that the effective pseudo spin for electron in N-layer graphene behaves like spin of fermionic- (bosonic-) particle when N is odd (even). The lattice wave states may be equivalent to the pseudo spin state of the form

$$\psi_{N,eff} = \begin{pmatrix} \psi_{A,1} \\ \psi_{B,N} \end{pmatrix} \to \begin{pmatrix} \psi_{+N/2} \\ \psi_{-N/2} \end{pmatrix},$$

where $\hat{S}_{z,N}\begin{pmatrix} \psi_{+N/2} \\ 0 \end{pmatrix} = +(N\hbar/2)\begin{pmatrix} \psi_{+N/2} \\ 0 \end{pmatrix}$ and $\hat{S}_{z,N}\begin{pmatrix} 0 \\ \psi_{-N/2} \end{pmatrix} = -(N\hbar/2)\begin{pmatrix} 0 \\ \psi_{-N/2} \end{pmatrix}$.

(10)

Unfortunately, the two-component pseudo spin state in eq.(10) may be usual only in case of N=1(monolayer), because there are two equivalent spin states $\psi_{A,1} \to \psi_{+1/2}$ and $\psi_{B,1} \to \psi_{-1/2}$ for spin of $\hbar/2$. In monolayer graphene, spin operators $\hat{S}_{x,1}, \hat{S}_{y,1}$ and $\hat{S}_{z,1}$ to do satisfy the "the commutation rule of an angular momentum", while for N>1, there are no two-component spin operators $\hat{S}_{x,N}$, $\hat{S}_{y,N}$ and $\hat{S}_{z,N}$ to satisfy such rule. If we define the spin operators

$$\hat{S}_{x,N} = N\frac{\hbar}{2}\sigma_x \text{ and } \hat{S}_{y,N} = N\frac{\hbar}{2}\sigma_y \tag{11}$$

for N-layer, we will get

$$\left[\hat{S}_{x,N}, \hat{S}_{y,N}\right] = iN\hbar\hat{S}_{z,N},\ \left[\hat{S}_{y,N}, \hat{S}_{z,N}\right] = iN\hbar\hat{S}_{x,N},\text{ and }\left[\hat{S}_{z,N}, \hat{S}_{x,N}\right] = iN\hbar\hat{S}_{y,N}. \tag{12}$$

This result in Eq.(12) violates "the commutation rule of an angular momentum", except for N=1. It is said that for N>1, pseudo spin operators defined in eq. (11) do not exhibit angular momentum operators. It would be better if we describe pseudo spin state in N-layer graphene as a real spin angular momentum state, because $\hat{S}_{z,N}$ should be a real angular momentum of Eigen spin $N\hbar/2$. This is related to the property that it obeys the conservation of total angular momentum $[\hat{H}_{N,eff}, \hat{J}_{z,N}] = 0$ [11, 12].

**4. New pseudo spin operator presentation**

Since pseudo spin operator given in eq.(9) violates "the commutation rule of an angular momentum" for N>1, despite it being an angular momentum in the z-direction. It must be wrong if it is an angular momentum, it does not obey the commutation rule of an angular momentum. In this work, we thus would like to propose new pseudo spin states $\tilde{\psi}_N$ for effective electron in N-layer graphene analogous to real spin states [19]. We may have

$$\tilde{\psi}_{N=even} = \left(\psi_{+N/2},\ \psi_{+(N-2)/2},\ \cdots,\ \psi_0,\ \cdots,\ \psi_{-(N-2)/2},\ \psi_{-N/2}\right)^T, \tag{13}$$

and

$$\tilde{\psi}_{N=odd} = \left( \psi_{+N/2}, \quad \psi_{+(N-2)/2}, \quad \cdots \quad \psi_{+1/2}, \quad \psi_{-1/2}, \quad \cdots \quad \psi_{-(N-2)/2}, \quad \psi_{-N/2} \right)^T, \tag{14}$$

The associated pseudo spin in the z-direction may be defined similar to real spin angular momentum operators [19], given by

$$\hat{\hat{S}}_{z,N=even} = \frac{\hbar}{2} \begin{pmatrix} +N & & & & & & 0 \\ & +(N-1) & & & & & \\ & & \ddots & & & & \\ & & & 0 & & & \\ & & & & \ddots & & \\ & & & & & -(N-1) & \\ 0 & & & & & & -N \end{pmatrix}, \tag{15}$$

and

$$\hat{\hat{S}}_{z,N=odd} = \frac{\hbar}{2} \begin{pmatrix} +N & & & & & & & 0 \\ & +(N-1) & & & & & & \\ & & \ddots & & & & & \\ & & & +1 & & & & \\ & & & & -1 & & & \\ & & & & & \ddots & & \\ & & & & & & -(N-1) & \\ 0 & & & & & & & -N \end{pmatrix}. \tag{16}$$

As we have seen from eq.(13) to eq.(16), the condition of $\psi_{A,1} = \psi_{+N/2}$ and $\psi_{B,N} = \psi_{-N/2}$ satisfy the Eigen value of spin $+N\hbar/2$ and $-N\hbar/2$, respectively. Also, $\{\psi_0, ..., \psi_{\pm(N-6)/2}, \psi_{\pm(N-4)/2}, \psi_{\pm(N-2)/2}\}$ and $\{\psi_{\pm 1/2}, ..., \psi_{\pm(N-6)/2}, \psi_{\pm(N-4)/2}, \psi_{\pm(N-2)/2}\}$ are introduced to be as Eigen states of $\hat{\hat{S}}_{z,N=even}$ and $\hat{\hat{S}}_{z,N=odd}$, respectively. We may set their amplitudes to be zero because there are no such states in multilayer graphene.

The Hamiltonian in eq.(8), may be replaced by a $(N+1)\times(N+1)$ matrix because it must act on $\tilde{\psi}_N$, given by

$$\hat{\hat{H}}_N = \begin{pmatrix} 0 & & & h_N \\ & & 0 & \\ & \cdot^{\cdot^{\cdot}} & & \\ & 0 & & \\ h_N^* & & & 0 \end{pmatrix},$$

where $h_N = \frac{(\hbar v_F)^N}{(-\gamma_\perp)^{N-1}}(\hat{k}_x - i\hat{k}_y)^N$.

(17)

Since we define the total angular momentum operator $\hat{\hat{J}}_{z,N} = \hat{L}_z + \hat{\hat{S}}_{z,N}$, the conservation of total angular momentum remains the same, ie.,

$$[\hat{\hat{H}}_N, \hat{L}_z + \hat{\hat{S}}_{z,N}] = 0. \qquad (18)$$

As mentioned above, the bosonic-like wave states $\{\psi_0, ..., \psi_{\pm(N-6)/2}, \psi_{\pm(N-4)/2}, \psi_{\pm(N-2)/2}\}$ and the fermionic-like wave state $\{\psi_{\pm 1/2}, ..., \psi_{\pm(N-6)/2}, \psi_{\pm(N-4)/2}, \psi_{\pm(N-2)/2}\}$ were introduced due to the fact that the Eigen states of associated with the revised pseudo spin operators in eqs.(15) and (16) may be required to preserve "the commutation rule of an angular momentum" for N>1, because spin operator must be $(N+1)\times(N+1)$ matrices. In the equation of motion, they would play no role, because there are no their matrix elements in the Hamiltonian $\hat{\hat{H}}_N$. Their motions may disappear in the Hamiltonian. The condition of which, the motions of $\psi_{A,1} = \psi_{+N/2}$ and $\psi_{B,N} = \psi_{-N/2}$ using the new Hamiltonian in eq.(17) remain the same as calculated via eq.(8), is also required.

From eqs.(14)- (18), we may obtain, for instance, result for pseudo wave states, pseudo spin operators and modified Hamiltonians which have been revised as

**N=1(spin-½ particle):**

Pseudo spin state: $\tilde{\psi}_1 = \begin{pmatrix} \psi_{+1/2} \\ \psi_{-1/2} \end{pmatrix}$, $\psi_{A,1} = \psi_{+1/2}$, $\psi_{B,1} = \psi_{-1/2}$

Spin operator: $\hat{\hat{S}}_{z,1} = \frac{\hbar}{2}\begin{pmatrix} 1 & 0 \\ 0 & -1 \end{pmatrix}$, $\hat{\hat{S}}_{x,1} = \frac{\hbar}{2}\begin{pmatrix} 0 & 1 \\ 1 & 0 \end{pmatrix}$, $\hat{\hat{S}}_{y,1} = \frac{\hbar}{2}\begin{pmatrix} 0 & -i \\ i & 0 \end{pmatrix}$

Hamiltonian: $\hat{\hat{H}}_1 = \begin{pmatrix} 0 & \hbar v_F(\hat{k}_x - i\hat{k}_y) \\ \hbar v_F(\hat{k}_x + i\hat{k}_y) & 0 \end{pmatrix}$

Conservation of total angular momentum $\hat{\hat{J}}_{z,1} = \hat{x}\hat{p}_y - \hat{y}\hat{p}_x + \hat{\hat{S}}_{z,1}$ : $[\hat{\hat{H}}_1, \hat{\hat{J}}_{z,1}] = 0$

(19)

**N=2 (spin-1 particle):**

Pseudo spin state: $\tilde{\psi}_2 = \begin{pmatrix} \psi_{+1} \\ \psi_0 \\ \psi_{-1} \end{pmatrix}$, $\psi_{A,1} = \psi_{+1}$, $\psi_{B,2} = \psi_{-1}$

Spin operator:

$$\hat{\hat{S}}_{z,2} = \hbar \begin{pmatrix} 1 & 0 & 0 \\ 0 & 0 & 0 \\ 0 & 0 & -1 \end{pmatrix}, \hat{\hat{S}}_{x,2} = \frac{\hbar}{\sqrt{2}} \begin{pmatrix} 0 & 1 & 0 \\ 1 & 0 & 1 \\ 0 & 1 & 0 \end{pmatrix}, \hat{\hat{S}}_{y,2} = \frac{\hbar}{\sqrt{2}} \begin{pmatrix} 0 & -i & 0 \\ i & 0 & -i \\ 0 & i & 0 \end{pmatrix}$$

Hamiltonian: $\hat{\hat{H}}_2 = \begin{pmatrix} 0 & 0 & \frac{(\hbar v_F)^2}{-\gamma_\perp}(\hat{k}_x - i\hat{k}_y)^2 \\ 0 & 0 & 0 \\ \frac{(\hbar v_F)^2}{-\gamma_\perp}(\hat{k}_x + i\hat{k}_y)^2 & 0 & 0 \end{pmatrix}$

Conservation of total angular momentum $\hat{\hat{J}}_{z,2} = \hat{x}\hat{p}_y - \hat{y}\hat{p}_x + \hat{\hat{S}}_{z,2}$ : $[\hat{\hat{H}}_2, \hat{\hat{J}}_{z,2}] = 0$

(20)

**N=3 (spin 3/2 particle):**

Pseudo spin state: $\tilde{\psi}_3 = \begin{pmatrix} \psi_{+3/2} \\ \psi_{+1/2} \\ \psi_{-1/2} \\ \psi_{-3/2} \end{pmatrix}$, $\psi_{A,1} = \psi_{+3/2}$, $\psi_{B,3} = \psi_{-3/2}$

Spin operator:

$$\hat{\hat{S}}_{z,3} = \frac{\hbar}{2} \begin{pmatrix} 3 & 0 & 0 & 0 \\ 0 & 1 & 0 & 0 \\ 0 & 0 & -1 & 0 \\ 0 & 0 & 0 & -3 \end{pmatrix}, \hat{\hat{S}}_{x,3} = \frac{\hbar}{2} \begin{pmatrix} 0 & \sqrt{3} & 0 & 0 \\ \sqrt{3} & 0 & 2 & 0 \\ 0 & 2 & 0 & \sqrt{3} \\ 0 & 0 & \sqrt{3} & 0 \end{pmatrix},$$

$$\hat{\hat{S}}_{y,3} = \frac{\hbar}{2}\begin{pmatrix} 0 & -i\sqrt{3} & 0 & 0 \\ i\sqrt{3} & 0 & -i2 & 0 \\ 0 & i2 & 0 & -i\sqrt{3} \\ 0 & 0 & i\sqrt{3} & 0 \end{pmatrix}$$

$$\text{Hamiltonian: } \hat{\hat{H}}_3 = \begin{pmatrix} 0 & 0 & 0 & \frac{(\hbar v_F)^3}{(-\gamma_\perp)^2}(\hat{k}_x - i\hat{k}_y)^3 \\ 0 & 0 & 0 & 0 \\ 0 & 0 & 0 & 0 \\ \frac{(\hbar v_F)^3}{(-\gamma_\perp)^2}(\hat{k}_x + i\hat{k}_y)^3 & 0 & 0 & 0 \end{pmatrix}$$

Conservation of total angular momentum $\hat{\hat{J}}_{z,3} = \hat{x}\hat{p}_y - \hat{y}\hat{p}_x + \hat{\hat{S}}_{z,3}$ : $[\hat{\hat{H}}_3, \hat{\hat{J}}_{z,3}] = 0$ .

(21)

All pseudo spin operators satisfy the commutation rule of an angular momentum to get

$$\left[\hat{\hat{S}}_{x,N}, \hat{\hat{S}}_{y,N}\right] = i\hbar\hat{\hat{S}}_{z,N}, \ \left[\hat{\hat{S}}_{y,N}, \hat{\hat{S}}_{z,N}\right] = i\hbar\hat{\hat{S}}_{x,N} \text{ and } \left[\hat{\hat{S}}_{z,N}, \hat{\hat{S}}_{x,N}\right] = i\hbar\hat{\hat{S}}_{y,N} .$$

(22)

## 5. Gap-induced pseudo Larmor precession for pseudo spin-½ and 1 particle

In this section, we would investigate pseudo Larmor precession of total angular momentum which is associated with pseudo spin $\hat{\hat{S}}_N = \hat{\hat{S}}_{x,N}\hat{i} + \hat{\hat{S}}_{y,N}\hat{j} + \hat{\hat{S}}_{z,N}\hat{k}$ in case of N=1 and N=2, in order to compare the Larmor frequency of pseudo spin ½ and spin 1 particles. The two-dimensional angular momentum operators must satisfy $\hat{L}_x = (\hat{y}\hat{p}_z - \hat{z}\hat{p}_y) = \hat{L}_y = (\hat{z}\hat{p}_x - \hat{x}\hat{p}_z) = 0$, to get

$$\hat{L} = \begin{pmatrix} \hat{y}\hat{p}_z - \hat{z}\hat{p}_y \\ \hat{z}\hat{p}_x - \hat{x}\hat{p}_z \\ \hat{x}\hat{p}_y - \hat{y}\hat{p}_x \end{pmatrix} = \begin{pmatrix} 0 \\ 0 \\ \hat{x}\hat{p}_y - \hat{y}\hat{p}_x \end{pmatrix}$$

(23)

Hence, the total angular momentum operator may be defined as

$$\hat{\tilde{J}}_N = \begin{pmatrix} \hat{\tilde{J}}_{x,N} \\ \hat{\tilde{J}}_{y,N} \\ \hat{\tilde{J}}_{z,N} \end{pmatrix} = \begin{pmatrix} \hat{\tilde{S}}_{x,N} \\ \hat{\tilde{S}}_{y,N} \\ (\hat{x}\hat{p}_y - \hat{y}\hat{p}_x) + \hat{\tilde{S}}_{z,N} \end{pmatrix}.$$

(24)

Torque of the total angular momentum operator may be determined using Heisenberg picture

$$\frac{d\hat{\tilde{J}}_N}{dt} = \frac{i}{\hbar}\left[\hat{\tilde{H}}_N, \hat{\tilde{J}}_N\right]. \quad (25)$$

We will firstly determine torque of pseudo spin-½ particle in monolayer graphene. The Hamiltonian for electron in monolayer graphene in case of gap opening may be usually given as [11, 12]

$$\hat{\tilde{H}}_1 = \begin{pmatrix} 0 & \hbar v_F(\hat{k}_x - i\hat{k}_y) \\ \hbar v_F(\hat{k}_x + i\hat{k}_y) & 0 \end{pmatrix} + \begin{pmatrix} \Delta & 0 \\ 0 & -\Delta \end{pmatrix}, \quad (26)$$

where $2\Delta$ is denoted as energy gap between valance and conduction bands. We consider the precession of the rest electron, satisfying $k_x = k_y = 0$.

$$\frac{d\hat{\tilde{J}}_1}{dt} = \frac{i}{\hbar}[\hat{\tilde{H}}_1, \hat{\tilde{J}}_1] = \begin{pmatrix} \frac{d\hat{\tilde{J}}_{x,1}}{dt} = -(\frac{2\Delta}{\hbar})\hat{\tilde{J}}_{y,1} \\ \frac{d\hat{\tilde{J}}_{y,1}}{dt} = +(\frac{2\Delta}{\hbar})\hat{\tilde{J}}_{x,1} \\ \frac{d\hat{\tilde{J}}_{z,1}}{dt} = 0 \end{pmatrix}. \quad (27)$$

The result above tells us that the frequency of the Larmor precession of the pseudo total angular momentum about the z-axis is $\omega_{Larmor} = 2\Delta/\hbar$ for electron in monolayer graphene.

We will firstly determine torque of spin-1 particle in bilayer graphene. The modified Hamiltonian for electron in bilayer graphene in case of gap opening may be given as

$$\hat{\hat{H}}_2 = \begin{pmatrix} 0 & 0 & \frac{(\hbar v_F)^2}{-\gamma_\perp}(\hat{k}_x - i\hat{k}_y)^2 \\ 0 & 0 & 0 \\ \frac{(\hbar v_F)^2}{-\gamma_\perp}(\hat{k}_x + i\hat{k}_y)^2 & 0 & 0 \end{pmatrix} + \begin{pmatrix} \Delta & 0 & 0 \\ 0 & 0 & 0 \\ 0 & 0 & -\Delta \end{pmatrix}. \quad (28)$$

When the precession of the rest electron is considered, we get

$$\frac{d\hat{\hat{J}}_2}{dt} = \frac{i}{\hbar}[\hat{\hat{H}}_2, \hat{\hat{J}}_2] = \begin{pmatrix} \frac{d\hat{\hat{J}}_{x,2}}{dt} = -(\frac{\Delta}{\hbar})\hat{\hat{J}}_{y,2} \\ \frac{d\hat{\hat{J}}_{y,2}}{dt} = +(\frac{\Delta}{\hbar})\hat{\hat{J}}_{x,2} \\ \frac{d\hat{\hat{J}}_{z,2}}{dt} = 0 \end{pmatrix}. \quad (29)$$

The Larmor precession of the pseudo total angular momentum about the z-axis is $\omega_{Larmor} = \Delta/\hbar$ for electron in bilayer graphene lower than that in monolayer graphene. It is said that both gaps in mono-and Bi-layer graphene would behave like pseudo magnetic field leading to the Larmor precessions. The magnetic-like interaction Hamiltonians in mono and bilayer graphene resulted from gap opening may be respectively written as

$$\hat{\hat{H}}_{1,int} = \begin{pmatrix} \Delta & 0 \\ 0 & -\Delta \end{pmatrix} = \mu_B \vec{B}_1.\hat{\hat{S}}_1 = \mu_B B_{z,1}\hat{\hat{S}}_{z,1},$$

and

$$\hat{\hat{H}}_{2,int} = \begin{pmatrix} \Delta & 0 & 0 \\ 0 & 0 & 0 \\ 0 & 0 & -\Delta \end{pmatrix} = \mu_B \vec{B}_2.\hat{\hat{S}}_2 = \mu_B \vec{B}_2 \hat{\hat{S}}_{z,2}, \quad (30)$$

where $\vec{B}_{1(2)} = (0)\hat{i} + (0)\hat{j} + B_{z,1(2)}\hat{k}$.

As we have seen in equation above, the gap-induced interacting magnetic strength in monolayer graphene is twice the magnitude in bilayer graphene ie., $\mu_B B_{z,1} = 2 \times \mu_B B_{z,2} = \Delta/\hbar$, where $\mu_B$ and $B_z$ are equivalent to Bohr magneton and magnetic field, respectively.

## 6. Summary and conclusion

In summary, we have studied the pseudo spin coupling to real orbital angular momentum in ABC-stacked N-layer graphene. The pseudo spin operator perpendicular to graphene sheet was calculated using the conservation of the total angular momentum of the z-direction [11, 12]. It is found that at low energy limit the pseudo spin of electron in N-layer ABC-stacked graphene does not exhibit spin-½ particle when N>1. Pseudo spin of electron is found to be of $N\times\hbar/2$. In this work, we have introduced new pseudo spin operators for electron in multilayer graphene satisfying "the commutation rule of an angular momentum". We have proposed graphene as a host material of electron with arbitrary pseudo spin, tunable by the number of layers.

## Acknowledgments

This work was supported by Kasetsart University Research and Development Institute (KURDI) and Thailand Research Fund (TRF) under Grant. No.TRG5780274.

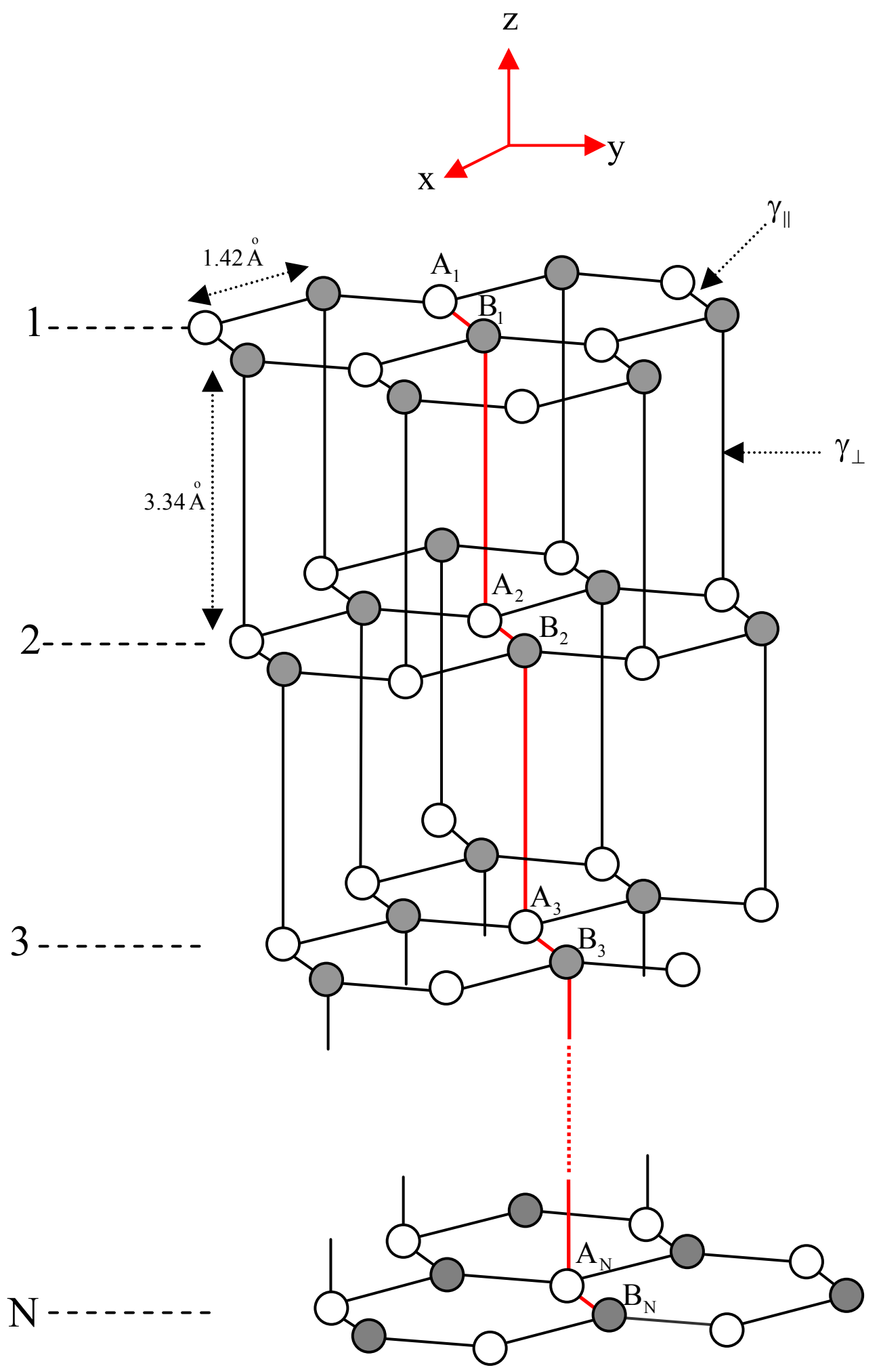

**Figure 1** Atomic structure of ABC-stacked N-layer graphene where $A_N$ and $B_N$ are denoted as the two sublatices in the $N^{th}$ -layer. The in-plane-nearest neighbor hoping energy $\gamma_{\parallel} \cong 2.8\,\text{eV}$ and the interlayer hoping energy $\gamma_{\perp} \cong 0.4$ eV are approximated [15].